\documentclass{AGUJournal}

\journalname{Geophysical Research Letters}

\begin{document}

\title{A climate network-based index to discriminate different types of El Ni\~no and La Ni\~na}

\authors{Marc Wiedermann\affil{1, 2}, Alexander
	Radebach\affil{3,4}, Jonathan F. Donges\affil{1,5}, J\"urgen
	Kurths\affil{1,2,6,7}, and Reik V. Donner\affil{1}}
\affiliation{1}{Potsdam Institute for Climate Impact Research ---
	Telegrafenberg A 31, 14412 Potsdam, Germany, EU}
\affiliation{2}{Department of Physics, Humboldt University --- Newtonstra\ss e 15,
	12489 Berlin, Germany, EU}
\affiliation{3}{Mercator Research Institute on Global Commons and Climate
Change --- Torgauer Stra\ss e 12-15, 10829 Berlin, Germany, EU}
\affiliation{4}{Economics of Climate Change, Technical University --- Stra\ss
	e des
	17. Juni 145, 10623 Berlin, Germany, EU}
\affiliation{5}{Stockholm Resilience Centre, Stockholm University --- Kr\"aftriket 2B, 114 19 Stockholm, Sweden, EU} 
\affiliation{6}{Institute for Complex Systems and Mathematical Biology,
  University of Aberdeen --- Aberdeen AB24 3FX, UK, EU}
\affiliation{7}{Department of Control Theory, Nizhny Novgorod State University ---
	Gagarin Avenue 23, 606950 Nizhny Novgorod, Russia}

\correspondingauthor{Marc Wiedermann}{marcwie@pik-potdam.de}


\begin{keypoints}
\item Discrimination between Eastern and Central Pacific El Ni\~no
\item New index based on climate networks for objective classification
\item Discriminations are possible for La Ni\~na as well
\end{keypoints}

\begin{abstract}
	El Ni\~no exhibits distinct Eastern Pacific (EP) and Central Pacific (CP)
	types which are commonly, but not always consistently, distinguished from
	each other by different signatures in equatorial climate variability.
	Here, we propose an index based on evolving climate networks to
	objectively discriminate between both flavors by utilizing a scalar-valued
	evolving climate network measure that quantifies spatial localization and dispersion in El
	Ni\~no's associated teleconnections. Our index displays a sharp peak (high
	localization) during EP events, whereas during CP events (larger dispersion)
	it remains close to the baseline observed during normal periods. In contrast
	to previous classification schemes, our approach specifically account for El Ni\~no's
	global impacts. We confirm recent El Ni\~no classifications for the years
	1951 to 2014 and assign types to those cases were former works
	yielded ambiguous results. Ultimately, we study La Ni\~na episodes and
	demonstrate that our index provides a similar discrimination into two types.
\end{abstract}

\section{Introduction}
The El Ni\~no Southern Oscillation (ENSO) alternates between positive (El
Ni\~no) and negative (La Ni\~na) phases~\citep{trenberth_definition_1997}. 
Especially the El Ni\~no phase further exhibits two distinct
types characterized by different spatial patterns of SST
anomalies~\citep[e.g.,][]{ashok_nino_2007,kao_contrasting_2009,kug_two_2009,yeh_nino_2009}.
The first type (the classic or Eastern Pacific (EP) El
Ni\~no~\citep{rasmusson_variations_1982,harrison_nino-southern_1998}) is
characterized by strong positive SST anomalies close to the western coast of
South America, while the second type (referred to as El Ni\~no Modoki or
Central Pacific (CP) El Ni\~no by different authors) exhibits the strongest SST
anomalies close to the dateline. Both types cause different impacts on the
global climate system, such as increased rainfall over north and eastern
Australia during CP El Ni\~nos~\citep{ashok_nino_2007,taschetto_nino_2009}
contrasted by a rainfall reduction over eastern Australia during EP El
Ni\~nos~\citep{chiew_nino/southern_1998}. Thus, a proper discrimination of
these types provides key information to assess El Ni\~no's possible impacts on
other climate subsystems.

While recent literature shows a large agreement on the classification of many
El Ni\~nos,
contradictory classifications arise in certain years such as, e.g.,
1986/1987, which has been classified as mixed~\citep{kug_two_2009},
EP~\citep{kim_unique_2011,yeh_nino_2009,hu_analysis_2011} or
CP~\citep{larkin_global_2005,hendon_prospects_2009,graf_central_2012}. In fact,
when reviewing existing studies~\citep{kim_impact_2009,
	kug_two_2009,kim_unique_2011,yeh_nino_2009,hu_analysis_2011,larkin_global_2005,hendon_prospects_2009,graf_central_2012},
8 out of 19 El Ni\~no events between 1957 and 2010 have not been classified in
agreement. 
These mismatches possibly arise since most discrimination schemes indeed
utilize the same climate observable (mostly SST), but apply different derived
characteristics such as the ENSO Modoki Index (EMI)~\citep{ashok_nino_2007}, 
the Nino3 and Nino4 index~\citep{kim_unique_2011,hu_analysis_2011}, or
empirical orthogonal function (EOF) analysis~\citep{kao_contrasting_2009,graf_central_2012}
to distinguish both El Ni\~no types. 

To provide a consistent and systematic discrimination, we propose here a method
to distinguish the two different El Ni\~no types based on the assessment of
time evolving complex climate networks~\citep{radebach_disentangling_2013}.
Climate networks consist of nodes representing time series and links displaying
some statistically relevant interdependency between
them~\citep{donges_complex_2009,tsonis_what_2006} ENSO has been studied
intensively using this tool to quantify corresponding
teleconnections~\citep{gozolchiani_emergence_2011,tsonis_topology_2008,tsonis_role_2008}
and its effect on other climatic
subsystems~\citep{gozolchiani_pattern_2008}. Recently, climate network
approaches allowed for successfully forecasting El Ni\~no by assessing the
strength of linkages in the equatorial
Pacific~\citep{ludescher_improved_2013,ludescher_very_2014}.

\citet{radebach_disentangling_2013} systematically studied the temporal
evolution of a global climate network in a spatially explicit way and linked
the resulting variability of its topology to the presence of the two different
El Ni\~no types. Following these lines of thought we develop a thorough
classification scheme that allows for an objective discrimination between EP
and CP El Ni\~nos. While most previous studies on El Ni\~no classification
focus on climate variability only within the equatorial Pacific, we
specifically acknowledge the global impact of ENSO\@. Our framework therefore
accounts for the correlation structure of global surface air temperature
anomalies (SATA), a variable that is highly affected by El
Ni\~no~\citep{yamasaki_climate_2008} and is, in contrast to SST, available
homogeneously sampled for the entire globe.

As an index that discriminates EP and CP El
Ni\~nos, we utilize the climate network's transitivity, a scalar-valued
measure, that quantifies the (disperse vs.\ strongly localized) spatial
distribution of pairwise correlations and teleconnections along the globe. 
First, we assess whether a certain period displays El Ni\~no conditions
according to the Oceanic Ni\~no Index (ONI). Second, we determine the
transitivity of evolving climate networks computed from one-year running window
cross-correlations 
with respect to a baseline value defined by the transitivity
of networks computed from 30-year windows that are centered around the period
of interest. 
The surpassing of that threshold defines an EP El Ni\~no, while
the contrary case indicates a CP El Ni\~no. In comparison with recent studies,
our methods confirms all EP and CP El Ni\~nos between 1951 and 2014 that were
commonly defined
by~\citet{kug_two_2009,kim_unique_2011,yeh_nino_2009,hu_analysis_2011,larkin_global_2005,hendon_prospects_2009}
and~\citet{graf_central_2012} and provides a consistent classification for
those periods that were ambiguously classified so far.

To consolidate our findings, we provide results for the climate network's node
strength fields during periods that our index defines as EP or CP
El Ni\~nos and show their similarity with patterns that are expected from an
EOF analysis~\citep{johnson_how_2013, donges_how_2015}. As recent
works~\citep{kug_are_2011,yuan_different_2012,tedeschi_influences_2013}
addressed the issue whether two types of La Ni\~na can be detected as well, we
perform the same procedure for these events and provide a similar 
discrimination for the negative phase of ENSO\@.

\section{Data}\label{sec:data}
We define El Ni\~no periods according to the Oceanic Ni\~no Index (ONI)
provided by the Climate Prediction Center of the National Oceanic and
Atmospheric Administration, which covers the time between
1950 and 2015 and is computed as the 3-month running mean SST anomaly in the
Nino3.4 region ($5^\circ$N-$5^\circ$S, $120^\circ$W-$170^\circ$W) with respect
to 
centered 30-year base periods that are updated every 5 years. As the
initial and final year of this data set include only incomplete information on
the 1951 La Ni\~na and the 2015 El Ni\~no, we restrict
ourselves to the period from 1951 to 2014.

We construct evolving climate networks from daily global surface air
temperature (SAT) data provided by the NCEP/NCAR reanalysis~\citep{kistler_ncepncar_2001} with a spatial resolution of $2.5^\circ$
in longitudinal and latitudinal direction covering the same time period as the
ONI\@. All 288 grid points located at the poles and all leap days are removed.
The data is anomalized in accordance with the definition of the ONI by
subtracting from the time series at every grid point the long-term annual cycle
computed over the same 30-year base periods as above that are updated every 5
years. Due to the lack of data before 1948 and after 2015, the years 1951 to
1965 are anomalized by the same base period (1951--1980) as the years 1965 to
1969. Similarly, the years 2005 to 2015 are anomalized by the 1986 to 2015 base
period. We acknowledge that this procedure induces small offsets in the time
series after every 5 years.  However, as we construct evolving climate networks
from time series of much shorter length we neglect these effects for the sake
of consistency with the definition of the ONI\@. The above anomalization
process ensures that once defined anomalies and ENSO periods are not altered by
the addition of more recent data.

Finally, we obtain $N=10,224$ time series $x_i(t)$ of surface air temperature
anomalies (SATA) with $N_t=23,360$ temporal sampling points each. 

\section{Methods}\label{sec:methods}
A climate network $G$ consists of a set of $N$ nodes that correspond to the
grid points in the underlying data set and a set of $M$ links which connect
pairs of nodes and indicate a strong statistical interrelationship between them.
The network is represented by its binary adjacency matrix $\textbf A$ with
entries $A_{ij}=1$ if two nodes $i$ and $j$ are linked and $A_{ij}=0$
otherwise~\citep{donges_backbone_2009,boers_complex_2013,stolbova_topology_2014}.
An extension of this procedure is the usage of an edge-weighted adjacency
matrix $\textbf W$ where $W_{ij}=0$ denotes the absence of a link, but
$W_{ij}>0$ denotes its strength (e.g., the pairwise correlation)~\citep{barrat_architecture_2004,hlinka_regional_2014,zemp_node-weighted_2014}.

\subsection{Network construction}
Following the framework of evolving climate network
analysis~\citep{radebach_disentangling_2013,hlinka_regional_2014} we construct
a sequence of networks $G_n$ from running-window cross-correlation matrices
$\textbf C_n=(C_{n,ij})$ between all pairs of SATA time series. A window $n$ is
characterized by its size $w$ and offset $d$ to the previous window. We choose
$d=30$ days and $w=365$ days to ensure that each window covers at least the
entire duration of an El Ni\~no or La Ni\~na episode. For each window $n$ we
obtain the truncated time series $\{x_{n,i}(t_n)\},\
t_n=\{nd, nd+1, \ldots, n+wd-1\}$ and compute the resulting $N_x\times N_x$
cross-correlation matrix $\textbf C_n$. In accordance with previous
studies~\citep{donges_complex_2009,tsonis_what_2006,palus_discerning_2011}, we
rely here on the linear Pearson correlation at zero lag. 

To reduce the complexity of $\textbf C_n$, it is advisable to represent only a
certain fraction $\rho$ of strongest absolute correlations as links between the
nodes~\citep{tsonis_what_2006,donges_complex_2009}. This yields an individual
threshold $T_n$ for each absolute correlation matrix $\textbf C_n^{abs} =
(|C_{n,ij}|)$ above which nodes are treated as linked. $\rho$ is then called
the \textit{link density} of $G_n$. Here, we keep $\rho=0.005$ fixed for all
windows $n$. This choice gives a number of $M$ links as low as possible to
ensure the consideration of only the strongest correlations. Further,
$\rho=0.005$ roughly corresponds to the fraction of nodes that
are situated inside the the Nino3.4 region.
We obtain thresholds (i.e., the lower bound of
absolute correlations values) $T_n$ in the range of 0.53 to 0.65. They are significant
above the $99\%$ significance level according to a standard student's t-test.

Binarizing $\textbf C_n^{abs}$ to an edge-unweighted adjacency matrix $\textbf A_n$
would neglect valuable information on the varying strength of
correlation between connected grid points. We therefore compute
edge-weighted adjacency matrices $\textbf W_n$ with entries $|C_{n,ij}|$ if two nodes
$i$ are $j$ are linked,
\begin{linenomath*}
	\begin{equation}
	W_{n,ij} = |C_{n,ij}| \cdot \Theta(|C_{n, ij}| - T_n).
\end{equation}
\end{linenomath*}
Due to the underlying grid type, the density of nodes increases
towards the poles inducing a systematic bias into
the computation of network measures~\citep{heitzig_node-weighted_2012}. 
This effect is corrected by assigning each node a weight $w_i$
corresponding to its latitudinal position
$\lambda_i$ on the grid~\citep{tsonis_what_2006,
	heitzig_node-weighted_2012,wiedermann_node-weighted_2013},
\begin{linenomath*}
	\begin{equation}
	w_i = \cos(\lambda_i).
\end{equation}
\end{linenomath*}
Network measures that include $w_i$ have been referred to as \textit{node
	splitting invariant} (n.s.i.)
measures~\citep{heitzig_node-weighted_2012,zemp_node-weighted_2014,wiedermann_node-weighted_2013}.

\subsection{Network transitivity}
El Ni\~no has a global impact on the climate system manifested by long-ranging
teleconnections into different regions of the
Earth~\citep{held_transients_1989,neelin_tropical_2003,trenberth_definition_1997}
which, in the context of climate networks can be regarded as mediators of
variations and fluctuations~\citep{tsonis_role_2008,runge_identifying_2015}.
Thus, El Ni\~no and its teleconnections cause a spatial organization
of high co-variability along the Earth's surface, which is reflected in 
the resulting climate network. The degree of
this organization can be quantified by a single-valued scalar metric,
the network
transitivity~\citep{watts_collective_1998,saramaki_generalizations_2007}, which
we use in its node-weighted form~\citep{heitzig_node-weighted_2012},
\begin{linenomath*}
	\begin{equation}
	\mathcal T_n = \frac{\sum_{i,j,k}w_iW_{n,ij}w_jW_{n,jk}w_kW_{n,ki}}
	{\sum_{i,j,k}w_iW_{n,ij}w_jW_{n,jk}w_k} \in [0,1].
\end{equation}
\end{linenomath*}
$\mathcal T_n$ gives the edge- and node-weighted fraction of closed triangles between
triples of nodes and measures how strongly the correlation in a
system under study or subsets thereof is spatially organized (high values) or
dispersed (low values). In a purely 
random network, $\mathcal T_n$ would naturally take very low values, i.e.,
approximately equal the link density in the standard case of no specific edge
and node-weights~\citep{erdos_evolution_1960}. $\mathcal T_n$ thus serves as a
good discriminator between phases of strong localization and high dispersion in
the global teleconnectivity of evolving climate
networks~\citep{radebach_disentangling_2013}. As EP and CP El Ni\~nos have
been shown to display different characteristics in their associated
teleconnections~\citep{ashok_nino_2007} we expect $\mathcal T_n$ to respond
differently to the presence of either of the two types. 

\subsection{Strength of individual nodes}
To connect our work with previous results from statistical climatology
we investigate for each node $i$ its corresponding area-weighted
strength 
\begin{linenomath*}
	\begin{equation}
	s_{n,i} = \sum_{j}w_jW_{n, ij}
\end{equation}
\end{linenomath*}
individually for each network $G_n$. $s_{n,i}$ measures the total weight of
links that are attached to each node $i$.  For the edge-unweighted case, this
measure reduces to the area-weighted connectivity~\citep{tsonis_role_2008}
which displays striking similarity with results from a node-weighted EOF
analysis~\citep{donges_how_2015,wiedermann_hierarchical_2015}. 

\section{Results}\label{sec:results}
The ONI identifies El Ni\~no (La Ni\~na) episodes if its values exceed (fall below) a
threshold of 0.5K (-0.5K) for at least 5 consecutive months,
yielding 22 (18) El Ni\~no (La Ni\~na)
episodes between 1951 and 2014 (Fig.~\ref{fig:transitivity}A). 

\subsection{Transitivity}
\begin{figure}[t]
\includegraphics[width=.9\linewidth]{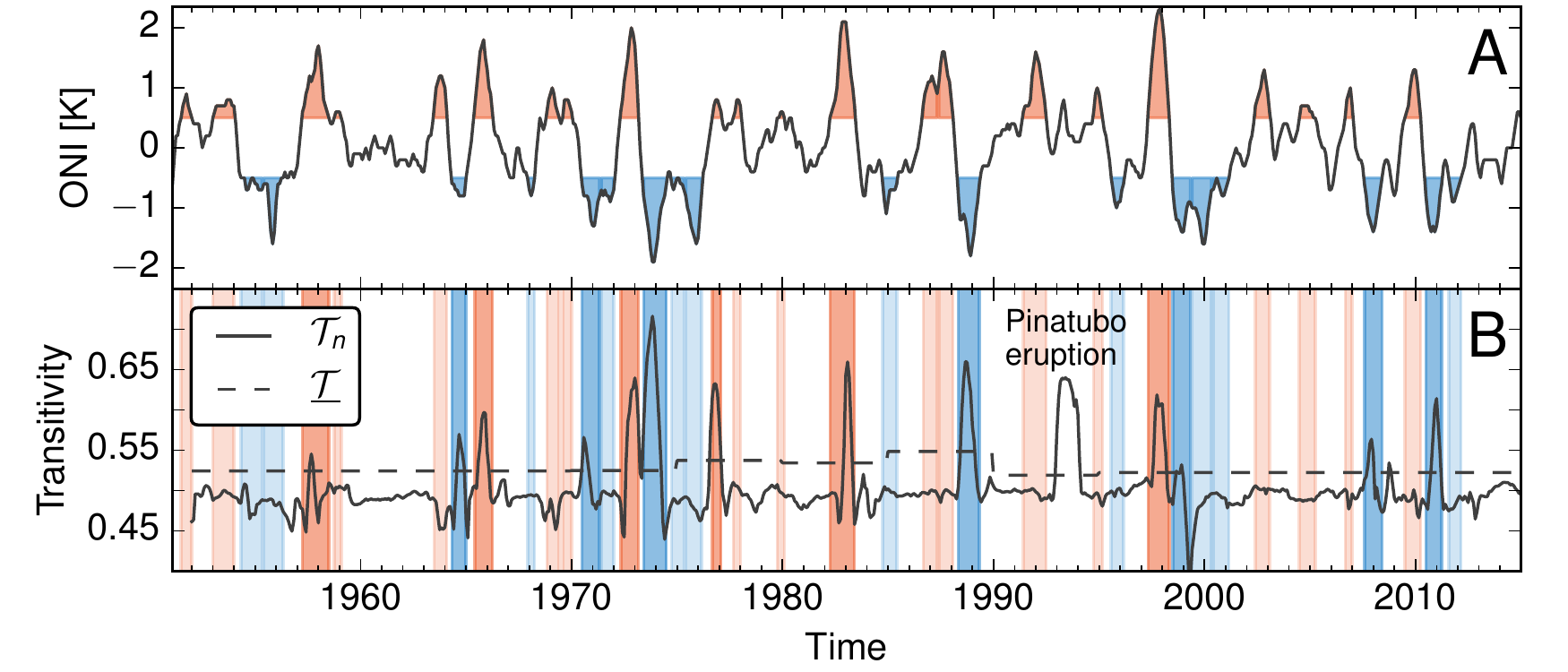}
\caption{(A) The ONI with El Ni\~no (La Ni\~na) periods marked in red (blue).
	(B) Time evolution of the evolving climate network's transitivity $\mathcal
	T_n$. The dashed vertical line indicates the baseline transitivity
	$\underline{\mathcal T_n}$.  Colored areas indicate El Ni\~no and La
	Ni\~na periods. Darker coloring indicates those periods where $\mathcal T_n$ exceeds $\underline{\mathcal T_n}$ and
	that are
	thus classified as EP type.}
\label{fig:transitivity}
\end{figure}

We construct $n=733$ evolving climate networks and compute their transitivity
$\mathcal T_n$ and node strength $s_{n,i}$. The end point of each window marks
the time at which the two measures are evaluated.
Figure~\ref{fig:transitivity}B shows the evolution of $\mathcal T_n$. Except
for one case with several 12-month time windows ending in 1993, which is likely
caused by the eruption of Mount Pinatubo in
1991~\citep{mccormick_atmospheric_1995,radebach_disentangling_2013}, distinct
peaks in $\mathcal T_n$ coincide exclusively with certain ENSO episodes. As
shown by~\citet{radebach_disentangling_2013}, the presence of an EP El Ni\~no
likely coincides with strong signals in (for their case unweighted)
transitivity, while no distinct signal is present during CP El Ni\~nos.
However, no quantitative criterion for this discrimination has been given so
far.

To give an objective definition of a \textit{strong} transitivity signal we
define a threshold value $\underline{\mathcal T}$ above which $\mathcal T_n$ is
considered to display a peak. We obtain an adaptive value of
$\underline{\mathcal T}$ as the transitivities of climate networks constructed
for the same 30-year periods that were used for the anomalization of the SAT
data and the derivation of the ONI\@. Thus, we compare all values of $\mathcal
T_n$ computed, e.g., during the period 1975--1979 with a baseline transitivity
$\underline{\mathcal T}$ computed for a climate network covering the 30-year
period of 1961--1990 (dashed line in Fig~\ref{fig:transitivity}B). This
procedure follows the definition of the ONI and we interpret
$\underline{\mathcal T}$ as representing the long-term average spatial organization in
the global climate network. Adaptively updating $\underline{\mathcal T}$ every
5 years automatically accounts for possible effects of long-term climate change
trends imprinting on the network statistics and the definition of
$\underline{\mathcal T}$ for periods in the past is not affected by the
addition of more recent data.

We detect six El Ni\~no periods during which $\mathcal T_n$ crosses
$\underline{\mathcal T}$ (dark red areas in Fig.~\ref{fig:transitivity}B)
corresponding to the El Ni\~nos of 1957, 1965, 1972, 1976, 1982, and 1997. For
all other El Ni\~nos $\mathcal T_n$ stays below $\underline{\mathcal T}$. In
the scope of our framework, we thus propose classifying the first case as EP
and the second case as CP events (light red areas in
Fig.~\ref{fig:transitivity}B).

\begin{table}[t]
	\caption{Recent classifications of El Ni\~no phases into CP and EP episodes. Asterisks
		denote mixed or undefined states. A hyphen denotes that no classification
		was performed for the specific year. Bold letters denote events where the
		network-based classification is in agreement with the reference. The last
		row summarizes the true positive rate (TPR) of our formalism. The
		second-last
		column indicates the largest overlap between all references and the last
		column summarizes the classification obtained from the network-based
		approach. 
	}\label{tab:overview}
	\resizebox{\textwidth}{!}{
\begin{tabular}{c|ccccccccc|c}\hline
          & Kug et al. & Kim et al. & Hu et al. & Larkin et al. & Hendon et al. & Graf et al. & Yeh et al. & Kim et al. & Common & Full\\
          & (2009)     & (2011)     & (2011)    & (2005)        & (2009)        & (2012)      & (2009)     & (2009)     & & results\\\hline
1953/1954 & -          & -          & -         & -             & -             & -           & -          & -          & -             & CP       \\
1957/1958 & -          & -          & \textbf{EP}       & \textbf{EP}           & -             & \textbf{EP}         & \textbf{EP}        & \textbf{EP}        & \textbf{EP}           & EP       \\
1958/1959 & -          & -          & -         & -             & -             & -           & -          & -          & -             & CP       \\
1963/1964 & -          & -          & -         & \textbf{CP}           & -             & \textbf{CP}         & EP         & EP         & -             & CP       \\
1965/1966 & -          & -          & \textbf{EP}       & \textbf{EP}           & -             & \textbf{EP}         & \textbf{EP}        & \textbf{EP}        & \textbf{EP}           & EP       \\
1968/1969 & -          & -          & \textbf{CP}       & \textbf{CP}           & -             & \textbf{CP}         & \textbf{CP}        & -          & \textbf{CP}           & CP       \\
1969/1970 & -          & -          & EP        & EP            & -             & -           & EP         & \textbf{CP}        & -             & CP       \\
1972/1973 & \textbf{EP}        & \textbf{EP}        & \textbf{EP}       & \textbf{EP}           & -             & \textbf{EP}         & \textbf{EP}        & \textbf{EP}        & \textbf{EP}           & EP       \\
1976/1977 & \textbf{EP}        & \textbf{EP}        & -         & \textbf{EP}           & -             & \textbf{EP}         & \textbf{EP}        & \textbf{EP}        & \textbf{EP}           & EP       \\
1977/1978 & \textbf{CP}        & \textbf{CP}        & -         & \textbf{CP}           & -             & \textbf{CP}         & \textbf{CP}        & -          & \textbf{CP}           & CP       \\
1979/1980 & -          & -          & -         & -             & -             & -           & *          & -          & -             & CP       \\
1982/1983 & \textbf{EP}        & \textbf{EP}        & \textbf{EP}       & \textbf{EP}           & \textbf{EP}           & \textbf{EP}         & \textbf{EP}        & \textbf{EP}        & \textbf{EP}           & EP       \\
1986/1987 & *          & EP         & EP        & \textbf{CP}           & \textbf{CP}           & \textbf{CP}         & EP         & -          & -             & CP       \\
1987/1988 & *          & -          & \textbf{CP}       & EP            & EP            & -           & EP         & EP         & -             & CP       \\
1991/1992 & *          & EP         & EP        & EP            & \textbf{CP}           & \textbf{CP}         & EP         & \textbf{CP}        & -             & CP       \\
1994/1995 & \textbf{CP}        & \textbf{CP}        & \textbf{CP}       & \textbf{CP}           & \textbf{CP}           & \textbf{CP}         & \textbf{CP}        & \textbf{CP}        & \textbf{CP}           & CP       \\
1997/1998 & \textbf{EP}        & \textbf{EP}        & \textbf{EP}       & \textbf{EP}           & \textbf{EP}           & \textbf{EP}         & \textbf{EP}        & \textbf{EP}        & \textbf{EP}           & EP       \\
2002/2003 & \textbf{CP}        & \textbf{CP}        & \textbf{CP}       & EP            & \textbf{CP}           & \textbf{CP}         & *          & \textbf{CP}        & -             & CP       \\
2004/2005 & \textbf{CP}        & \textbf{CP}        & -         & -             & \textbf{CP}           & \textbf{CP}         & \textbf{CP}        & \textbf{CP}        & \textbf{CP}           & CP       \\
2006/2007 & -          & EP         & \textbf{CP}       & \textbf{CP}           & -             & -           & EP         & -          & -             & CP       \\
2009/2010 & -          & \textbf{CP}        & -         & -             & -             & \textbf{CP}         & -          & -          & \textbf{CP}           & CP       \\\hline
TPR       & 1.0        & 0.57       & 0.62      & 0.6           & 0.67          & 1.0         & 0.5        & 0.75       & 1.0          \\
\end{tabular}
}
\end{table}
For comparison, the proposed classifications of El Ni\~no phases into
EP and CP types from eight recent studies~\citep{kim_impact_2009,
	kug_two_2009,kim_unique_2011,yeh_nino_2009,hu_analysis_2011,larkin_global_2005,hendon_prospects_2009,graf_central_2012}
 are summarized in Tab.~\ref{tab:overview}.
 To quantify the consistency of the network-based discrimination, we define a true
positive rate (TPR) as the fraction of EP El Ni\~nos in each study that are
detected by our framework. Accordingly, the false positive rate (FPR) is the
fraction of CP El Ni\~nos in each study that our method classifies as EP type.
With respect to all references we obtain a FPR of zero. The TPR for each
reference is presented in the last row of Tab.~\ref{tab:overview}. Its values
vary between 1 for the comparison with~\citet{graf_central_2012}
and~\citet{hu_analysis_2011}, and 0.5 for the comparison
with~\citet{yeh_nino_2009}. Furthermore, we note that among all references 8
out of 19 events are not classified in agreement. Taking only the mutual
agreement between all references as a basis for testing, we confirm all past
classifications (second-last column in Tab.~\ref{tab:overview}). To provide
results for the eight ambiguously defined periods, the network-based
classification for all El Ni\~nos is given in the last column of
Tab.~\ref{tab:overview}.

We find the largest consistency with the results from \citet{graf_central_2012}
which are obtained from an EOF analysis, a framework that, like our method, is
based on the evaluation of cross-correlations between different grid points.
This methodological congruence may explain the good agreement between the
results and confirms the validity of our work.  However, the advantage of
utilizing a network-based approach instead of EOFs is that the entire spatial
structure of the underlying covariance patterns is reduced to a single index.
Moreover, its evaluation does not rely on any visual inspection, but provides
an objective binary classification depending on whether or not the short-term
transitivity $\mathcal T_n$ exceeds its long-term baseline $\underline{\mathcal
	T}$.

We repeat the analysis for La Ni\~na periods and classify 7 EP (1964, 1970,
1973, 1988, 1998, 2007, 2010) and 11 CP (1954, 1955, 1967, 1971, 1974, 1975,
1984, 1995, 2000, 2001, and 2011) periods (dark (EP) and light (CP) blue areas
in Fig.~\ref{fig:transitivity}B).  Even though references providing actual
discriminations of the different La Ni\~na years are scarse, we compiled two
recent works and confirm the reported EP La Ni\~nas of 1964 and
1970~\citep{yuan_different_2012} and CP La Ni\~nas of 1975, 1984, 2000, 2001,
and 2011~\citep{yuan_different_2012,tedeschi_influences_2013}.  
Future work should further evaluate the discrimination of La Ni\~na periods
proposed by our method. 
	
\subsection{Node strength}
\begin{figure}[t]
\includegraphics[width=.9\linewidth]{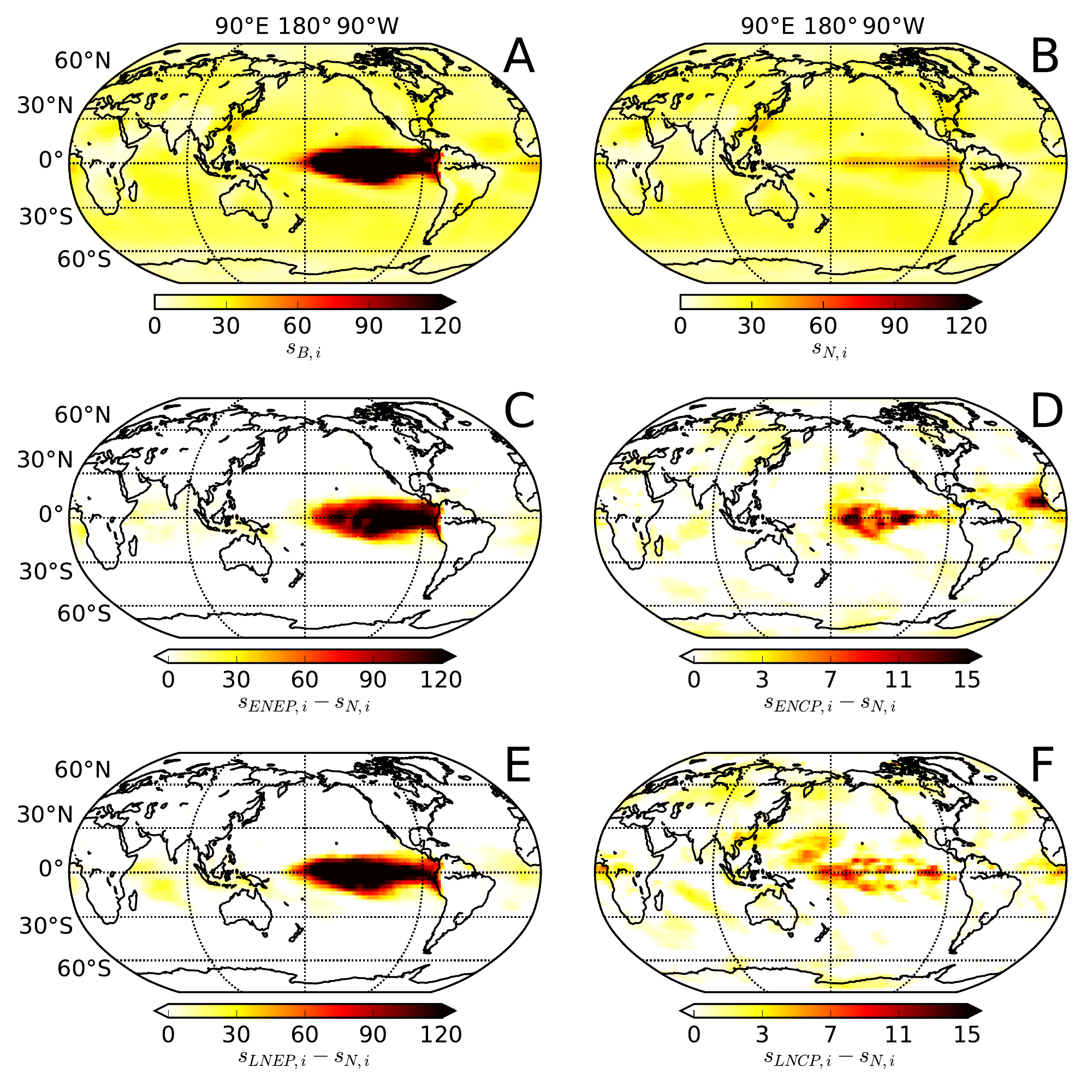}
\caption{(A) 
	Average strength of nodes in the baseline climate networks. (B) Average node strength
	of the evolving networks during normal periods.
	(C) Differences between the average
	node strength during El Ni\~no periods that are classified as EP type
	and the average node strength during normal periods. (D) The same as in (C)
	for El Ni\~no periods that are classified as CP type. (E,F) The same as in
	(C,D) for La Ni\~na periods.}
\label{fig:degree}
\end{figure}

To further consolidate our findings we compute the average node strengths $s_{B,i}$
from the six networks that are used to define $\underline{\mathcal T}$
(Fig.~\ref{fig:degree}A). We obtain the highest values in the equatorial
Pacific highlighting ENSO's importance in the global climate network.
Additionally, we compute the average node strength $s_{N, i}$ taken over all
\textit{normal} periods, i.e., those periods where neither El Ni\~no or La
Ni\~na are present (Fig.~\ref{fig:degree}A). As by its definition the effect of
ENSO is reduced, $s_{N, i}$ displays comparably low values and a more
homogeneous distribution across the entire globe than $s_{B,i}$. Ultimately, we
calculate the average node strength $s_{ENEP,i}$ ($s_{ENCP,i}$) taken over all
El Ni\~no periods that our method classifies as EP (CP) type (see also
Fig.~\ref{fig:transitivity}B). To investigate the deviation from the normal
state during either of the two periods we display their differences from $s_{N,
	i}$ in Fig.~\ref{fig:degree}C,D. For EP El Ni\~nos (Fig.~\ref{fig:degree}C)
we find an expected maximum in the equatorial Pacific, which is the typical
ENSO-related pattern known from a classical EOF
analysis~\citep{johnson_how_2013}. For CP El Ni\~nos we find a weakening of
this pattern and a westward shift of the maxima towards the dateline.
This pattern has been observed in the corresponding EOFs as
well~\citep{johnson_how_2013}. However, we note that $s_{ENCP,i}$ only differs
from $s_{N, i}$ to a small amount (Fig.~\ref{fig:degree}D). This again
suggests, that during CP El Ni\~nos the evolving climate networks exhibit a
similar state as during normal periods. We compute similar average quantities,
$s_{LNEP,i}$ and $s_{LNEP,i}$, for La Ni\~na events and again evaluate their
deviations from the normal state (Fig.~\ref{fig:degree}E,F). We find
quantitatively and qualitatively similar patterns as for El Ni\~no, which
highlights the symmetry in the statistics of the two ENSO phases. Even though a
similarly thorough comparison with existing literature is not yet possible for
La Ni\~na, the high congruence between $s_{ENEP,i}$ and $s_{LNEP,i}$
($s_{ENCP,i}$ and $s_{LNCP,i}$) suggests that our discrimination scheme
provides reasonable results for La Ni\~na phases as well. 

\subsection{Robustness}
\begin{figure}[t]
\includegraphics[width=.9\linewidth]{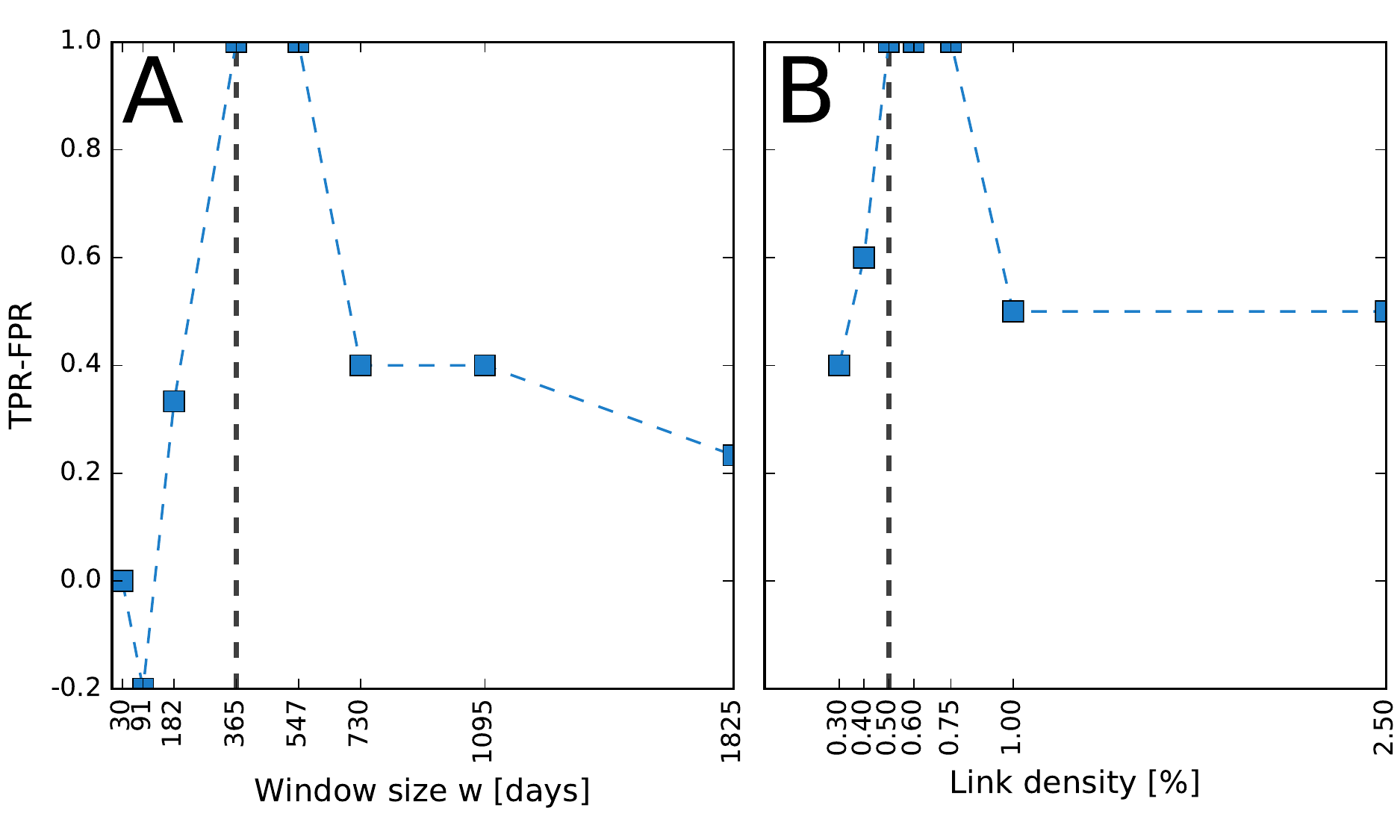}
\caption{(A) Difference between true positive rate (TPR) and false positive
	rate (FPR) for classifications obtained from the network approach and the
	largest overlap between all references in Tab.~\ref{tab:overview} for
	different window sizes $w$ and fixed link density $\rho=0.005$. (B) The
	same as in (A) for different link densities $\rho$ and fixed window
	size $w=365$ days. Dashed lines indicate the choices of
	parameters that yield the results in Fig.~\ref{fig:transitivity}B and
	Fig.~\ref{fig:degree}.  }
\label{fig:robustness}
\end{figure}

To evaluate the robustness of our results with respect to the window size $w$
and link density $\rho$, we vary both parameters individually and assess the
difference between the TPR and FPR when testing our classification against the
largest overlap of the literature (second-last column in
Tab.~\ref{tab:overview}). This score takes its maximum value of $1$ if our
method confirms the literature's classification of each event and is lower
otherwise. Figure~\ref{fig:robustness}A (Fig.~\ref{fig:robustness}B) shows the
score for different $w$ ($\rho$) and fixed $\rho=0.005$ ($w=365$ days). The
highest scores are obtained for window sizes between $w=365$ and $w=547$ days
and link densities between $\rho=0.005$ and $\rho=0.0075$. Shorter window sizes
cause a reduction of the score as the windows become too small to sufficiently
cover the temporal extent of an ENSO event. For larger window sizes the effect
of ENSO is suppressed by including too many of the normal periods into each
window. The link density of $\rho=0.005$ was initially chosen as it roughly
corresponds to the fraction of nodes located inside the Nino3.4 region. Smaller
values cause the network to be only composed of highly correlated trivial
nearest-neighbor connections and teleconnections with comparably lower pairwise
cross-correlation vales are not captured. In contrast, larger values result in
too many trivial links alongside those attributed to the effects of ENSO\@.
Generally, the score varies smoothly along the range of parameters and shows
maximum values for our initial choices. Thus, we consider our results to be
sufficiently robust.

\section{Conclusion}\label{sec:conclusion}
We have proposed an index based on evolving climate networks to objectively
discriminate between EP and CP types of El Ni\~no and La Ni\~na episodes. It
relies on the evolution of the networks' transitivity, measuring spatial
localization and dispersion of strong cross-correlations between different grid
points in a global SATA field.  If this index peaks during an ENSO phase, it
detects the presence of an EP type event. In contrast, the absence of a
remarkable signal during an ENSO period indicates CP type events.  From the
climate network perspective this indicates an increased localization and
clustering of teleconnections during EP phases in comparison with CP and normal
phases where teleconnections seem to appear more dispersed.  Our method does
not require any visual inspection or manual thresholding of observed patterns
but objectively categorizes ENSO phases into different types by intercomparing
the networks' short-term ($\mathcal T_n$) and long-term states
($\underline{\mathcal T}$).

In comparison with eight recent works on El Ni\~no classification our method
confirms the classification of years that all references have in
common and provides a discrimination for those years that where so far
ambiguously defined. Compared to approaches based on average temperature
observations our method produces a sharp signal in the variable under study,
i.e, the network transitivity, and thus provides a clear distinction between
the two types of El Ni\~no episodes.

Even though references are scarce, our findings also confirm different recently
reported EP and CP La Ni\~na periods and show that our discrimination scheme is
applicable to this negative phase of ENSO as well.

Future work should investigate more thoroughly the spatial distribution of
links in the evolving climate networks during different
ENSO stages to gain a more systematic understanding of the physical
mechanisms behind the observed differences in transitivity. 
Moreover, being automated and objective, our framework allows for a
systematic evaluation of climate model simulations and could be used to
investigate potential changes in the projected frequency of the two ENSO
flavors in the future, e.g., due to anthropogenic global
warming~\citep{yeh_nino_2009}.

\acknowledgments
	MW and RVD have been supported by the German Federal Ministry for Education
	and Research via the BMBF Young Investigators Group CoSy-CC$^2$ (grant no.
	01LN1306A). JFD thanks the Stordalen Foundation via the Planetary Boundary
	Research Network (PB.net) and the Earth League's EarthDoc programme for
	financial support. JK acknowledges the IRTG 1740 funded by DFG and FAPESP\@.
	NCEP Reanalysis data is provided by the NOAA/OAR/ESRL PSD, Boulder, Colorado,
	USA, from their website \texttt{http://www.esrl.noaa.gov/psd/}.
	Parts of the analysis have been performed using the Python package
	\texttt{pyunicorn}~\citep{donges_unified_2015} available at
	\texttt{https://github.com/pik-copan/pyunicorn}.




\listofchanges

\end{document}